\documentclass{article}    
\usepackage{latexsym}
\usepackage{amsmath,amsthm,amsfonts}
\usepackage{amssymb}
\usepackage{graphicx}
\usepackage{float}

\newcommand{\VaR}{ \mbox{VaR} }
\newcommand{\ES}{ \mbox{ES} }
\newcommand{\Ex}{ \mathbb{E} }
\newcommand{\Pb}{ \mathbb{P} }

\newtheorem{theorem}{Theorem}

\newtheorem{proposition}[theorem]{Proposition}
\newtheorem{remark}[theorem]{Remark}
\newtheorem{assumption}[theorem]{Assumption}

\title{ \bf Liquidity-adjusted Market Risk Measures with Stochastic Holding Period}  
\author{ Damiano Brigo \\ Dept. of
Mathematics \\ King's College, London \\ {\tt damiano.brigo@kcl.ac.uk} \and Claudio Nordio \\ Risk Management \\ Banco Popolare, Milan \\ {\tt claudio.nordio@bancopopolare.it} \footnote{Preliminary version. Comments welcome. This paper reflects the authors' opinions and
not necessarily those of their current and past employers. The authors are grateful to Dirk Tasche for helpful suggestions and correspondence}}
\date{First version: Sept 20, 2010. This version: \today}

\begin{document}           
\maketitle                 

\pagestyle{myheadings}
\markboth{}{D. Brigo, C. Nordio: Liquidity-adjusted Market Risk Measures with Stochastic Holding Period}



\centerline{\bf \large Abstract}

\begin{quotation} \em \small
Within the context of risk integration, we introduce in risk
measurement stochastic holding period (SHP) models. This is done in
order to obtain a `liquidity-adjusted risk measure' characterized by
the absence of a fixed time horizon. The underlying assumption is
that - due to changes on market liquidity conditions - one operates
along an `operational time' to which the P\&L process of liquidating
a market portfolio is referred. This framework leads to a
mixture of distributions for the portfolio returns, potentially allowing for
skewness, heavy tails and extreme scenarios. We analyze the impact of possible
distributional choices for the SHP. In a multivariate setting, we
hint at the possible introduction of dependent SHP processes, which
potentially lead to non linear dependence among the P\&L processes
and therefore to tail dependence across assets in the
portfolio, although this may require drastic choices on the SHP
distributions. We also find that increasing dependence
as measured by Kendall's tau through common SHP's appears to be unfeasible.
We finally discuss potential developments following
future availability of market data.
\end{quotation}

\bigskip

{\bf JEL} Classification codes: C15, C16, G10, G18

{\bf AMS} Classification codes: 62G32, 91B28, 91B70

\bigskip


\vspace{5mm}

\section{Introduction}

According to the IMCR research group of BCBS \cite{fi}, {\it ``liquidity conditions interact with market risk and credit risk through the horizon over which assets can be liquidated''}. To face the impact of market liquidity risk, risk managers agree in adopting a longer holding period to calculate the market VaR, for instance ten business days instead of one; recently, BCBS has prudentially stretched such liquidity horizon to three months \cite{gc}. However, even the IMCR group pointed out that {\it ``the liquidity of traded products can vary substantially over time and in unpredictable ways''}, and moreover, {\it ``IMCR studies suggest that banks' exposures to market risk and credit risk vary with liquidity conditions in the market''}. The former statement suggests a stochastic description of the time horizon over which a portfolio can be liquidated, and the latter highlights a dependence issue.
\\\newline
We can start by saying that probably the holding period of a risky portfolio is neither ten business days nor three months; it could, for instance, be 10 business days with probability 99\% and three months with probability 1\%. This is a very simple assumption but it may have already interesting consequences. Indeed, given the FSA requirement to justify liquidity horizon assumptions for the Incremental Risk Charge modelling, a simple example with the two-points liquidity horizon distribution that we develop below  could be interpreted as a mixture of the distribution under normal conditions and of the distribution under stressed and rare conditions.

To make the general idea more precise, it is necessary to distinguish between the two processes:
\begin{itemize}
  \item the daily P\&L of the risky portfolio;
  \item the P\&L of disinvesting and reinvesting in the risky portfolio.
\end{itemize}
In the following we will assume no transaction costs, in order to fully represent the liquidity risk through the holding period variability. Therefore, even if the cumulative P\&L is the same for the two processes above on the long term, the latter has more variability than the former, due to variable liquidity conditions in the market. If we introduce a third process, describing the dynamics of such liquidity conditions, for instance
\begin{itemize}
  \item the process of time horizons over which the risky portfolio can be fully bought or liquidated
\end{itemize}
then the P\&L is better defined by the returns calculated over such stochastic time horizons instead of a daily basis. We will use the ``stochastic holding period'' (SHP) acronym for that process, which belongs to the class of {\bf positive processes} largely used in mathematical finance.
We define the liquidity-adjusted VaR or Expexted Shortfall (ES) of a risky portfolio as the VaR or ES of portfolio returns calculated over the horizon defined  by the SHP process, which is the `operational time' along which the portfolio manager must operate, in contrast to the `calendar time' over which the risk manager usually measures VaR.
\\ \newline

Earlier literature on extending risk measures to liquidity includes several studies. Jarrow and Subramanian (1997), Bangia et al. (1999), Angelidis and Benos (2005), Jarrow and Protter (2005), Stange and Kaserer (2008), Earnst, Stange and Kaserer (2009), among few others, propose different methods of extending risk measures to account for liquidity risk. Bangia et al. (1999) classify market liquidity risk in two categories: (a) the exogenous illiquidity which depends on general market conditions, is common to all market players and is unaffected by the actions of any one participant and (b) the endogenous illiquidity that is specific to one's position in the market, varies across different market players and is mainly related to the impact of the trade size on the bid-ask spread.  Bangia et al. (1999) and Earnst et al. (2009) only consider the exogenous illiquidity risk and propose a liquidity adjusted VaR measure built using the distribution of the bid-ask spreads. The other mentioned studies model and account for endogenous risk in the calculation of liquidity adjusted risk measures.
In the context of the coherent risk measures literature, the general axioms  a liquidity measure should satisfy are discussed in \cite{acerbi}.

None of the above works however focuses specifically on our setup with random holding period, which represents a simple but powerful idea to include liquidity in traditional risk measures such as Value at Risk or Expected Shortfall. When analyzing multiple positions, holding periods can be taken to be strongly dependent, in line with the first classification (a) of Bangia et al (1999) above, or independent, so as to fit the second category (b). We will discuss whether adding dependent holding periods to different positions can actually add dependence to the position returns.
\\ \newline

The paper is organized as follows. In order to illustrate the SHP model, first in a univariate case (Section 2) and then in a bivariate one (Section 3), it is considerably easier to focus on examples on (log)normal processes. A brief colloquial hint at positive processes is presented in Section 2, to deepen the intuition of the impact on risk measures of introducing a SHP process. Across Section 3 and Section 4, where we try to address the issue of calibration, we outline a possible multivariate model which could be adopted, in line of principle, in a top-down approach to risk integration in order to include the liquidity risk and its dependence on other risks.

Finally, we point out that this paper is meant as a proposal to open a research effort in stochastic holding period models for risk measures. This paper contains several suggestions on future developments, depending on an increased availability of market data. The core ideas on the SHP framework, however, are presented in this opening paper.

\section{The univariate case}

Let us suppose we have to calculate VaR of a market portfolio whose value at time $t$ is $V_t$. We call $X_t = \ln V_t$, so that the log-return on the portfolio value at time $t$ over a period $h$ is
\[ X_{t + h} - X_t = \ln(V_{t + h}/V_t) \approx \frac{V_{t+h}-V_t}{V_t} .  \]
In order to include liquidity risk, the risk manager decides that a realistic, simplified statistics of the holding period in the future will be
\begin{table}[H]
\caption{Simplified discrete SHP}
\centering
\begin{tabular}{c c}
  \hline\hline
  Holding Period & Probability \\[0.5ex]
  \hline
  10 & 0.99 \\
  75 & 0.01 \\[1ex]
  \hline
\end{tabular}
\label{shp_bernoulli}
\end{table}
To estimate liquidity-adjusted VaR say at time 0, the risk manager will perform a number of simulations of  $V_{0+H_0}-V_{0}$ with $H_0$ randomly chosen by the statistics above, and finally will calculate the desired risk measure from the resulting distribution. If the log-return $X_{T}-X_{0}$ is normally distributed with zero mean and variance $T$ for deterministic $T$ (e.g. a Brownian motion, i.e. a Random walk), then the risk manager could simplify the simulation using $X_{0+H_0}-X_{0}|_{H_0}\stackrel{d}\backsim\sqrt{H_0}\left(X_{1}-X_{0}\right)$ where $|_{H_0}$ denotes ``conditional on $H_0$".
With this practical exercise in mind, let us generalize this example to a generic $t$.

\subsection{A brief review on the stochastic holding period framework}


A process for the risk horizon at time t, i.e. $t \mapsto H_t$, is a positive stochastic process modeling the risk horizon over time. We have that the risk measure at time $t$ will be taken on the change in value of the portfolio over this random horizon. If $X_t$ is the log-value of the portfolio at time $t$, we have that the risk measure at time $t$ is to be taken on the log-return
\[ X_{t+H_t} - X_t .\]
For example, if one uses a $99\%$ Value at Risk (VaR) measure, this will be the 1st percentile of
$X_{t+H_t} - X_t$. The request that $H_t$ be just positive means that the horizon at future times can both increase and decrease, meaning that liquidity can vary in both directions.

There is a large number of choices for positive processes: one can take lognormal processes with or without mean reversion, mean reverting square root processes, squared gaussian processes, all with or without jumps. This allows one to model the holding period dynamics as mean reverting or not, continuous or with jumps, and with thinner or fatter tails. Other examples are possible, such as Variance Gamma or mixture processes, or Levy processes. See for example \cite{brigotoolkit1} and \cite{brigotoolkit2}.

\subsection{Semi-analytic Solutions and Simulations}

Going back to the previous example, let us suppose that

\begin{assumption}\label{ass:dist} The increments $X_{t+1y}-X_t$ are logarithmic returns of an equity index, normally distributed with annual mean and standard deviation respectively $\mu_{1y} = -1.5\%$ and $\sigma_{1y} =30\%$.
\end{assumption}

We suppose an exposure of 100 in domestic currency.

Before running the simulation, we recall some basic notation and formulas.

The portfolio log-returns under random holding period at time $0$ can be written as

\[
P[X_{H_0} - X_0 < x] = \int_0^\infty
P[X_{h} - X_0 < x] d F_{H,t}(h)
 \]

i.e. as a mixture of Gaussian returns, weighted by the holding period distribution. Here $F_{H,t}$ denotes the cumulative distribution function of the holding period at time $t$, i.e. of $H_t$.

\begin{remark}{\bf (Mixtures for heavy-tailed and skewed distributions).}
Mixtures of distributions have been used for a long time in statistics and may lead to heavy tails, allowing for modeling of skewed distributions and of extreme events. Given the fact that mixtures lead, in the distributions space, to linear (convex) combinations of possibly simple and well understood distributions, they are tractable and easy to interpret. The literature on mixtures is enormous and it is impossible to do justice to all this literature here. We just hint at the fact that
static mixtures of distributions had been postulated in the past to fit option prices for a given maturity, see for example \cite{ritchey}, where a mixture of normal
densities for the density of the asset log-returns under the pricing measure is assumed, and subsequently \cite{melick}, \cite{bhupinder},
and \cite{guo}.  In the last decade \cite{brigobachelier}, \cite{brigorisk} and \cite{alexander} have extended the mixture distributions to fully dynamic arbitrage-free stochastic processes for asset prices.
\end{remark}

Going back to our notation, $\VaR_{t,h,c}$ and $\ES_{t,h,c}$ are the value at risk and expected shortfall, respectively, for an horizon $h$ at confidence level $c$ at time $t$, namely

\[ \Pb\{X_{t+h} - X_t > - \VaR_{t,h,c} \} = c, \ \ \
\ES_{t,h,c} =  - \Ex[ X_{t+h} - X_t | X_{t+h} - X_t \le - \VaR_{t,h,c}]. \]


In the gaussian log-returns case where

\begin{equation}\label{eq:normaldX} X_{t+h} - X_t \ \ \mbox{is normally distributed with mean}\ \ \mu_{t,h}
\ \  \mbox{and standard deviation} \ \  \sigma_{t,h}
\end{equation}

we get

\[ \VaR_{t,h,c} = -\mu_{t,h} + \Phi^{-1}(c) \sigma_{t,h}, \ \
\ES_{t,h,c} = -\mu_{t,h} + \sigma_{t,h} p(  \Phi^{-1}(c) )/(1-c)   \]

where $p$ is the standard normal probability density function and $\Phi$ is the standard normal cumulative distribution function.

In the following we will calculate VaR and Expected Shortfall referred to a confidence level of $99.96\%$, calculated over the fixed time horizons of 10 and 75 business days, and under SHP process with statistics given by Table \ref{shp_bernoulli}, using Monte Carlo simulations. Each year has 250 (working) days.

\begin{table}[H]
\caption{SHP distributions and Market Risk}
\centering
\begin{tabular}{c c c c c}
  \hline\hline
  Holding Period & VaR 99.96\% & (analytic) & ES 99.96\% & (analytic) \\[0.5ex]
  \hline
  constant 10 b.d. & 20.1 & (20.18) &  21.7 & (21.74)\\
  constant 75 b.d. & 55.7 & (55.54) &  60.0 & (59.81) \\
  SHP (Bernoulli 10/75, $p_{10}$=0.99) & 29.6 & (29.23) & 36.1 & (35.47) \\[1ex]
  \hline
\end{tabular}\label{fmr_bernoulli}
\end{table}

More generally, we may derive the VaR and ES formulas for the case where $H_t$ is distributed according to a general distribution

\[ \Pb( H_t \le x ) = F_{H,t}(x), \ \ x \ge 0  \]

and

\[ \Pb( X_{t+h} - X_t \le x ) = F_{X,t,h}(x) .  \]

We define VaR and ES under a random horizon $H_t$ at time $t$ and for a confidence level $c$ as
\[ \Pb\{X_{t+H_t} - X_t > - \VaR_{H,t,c} \} = c, \ \ \
\ES_{H,t,c} =  - \Ex[ X_{t+H_t} - X_t | X_{t+H_t} - X_t \le - \VaR_{H,t,c}]. \]

Using the tower property of conditional expectation it is immediate to prove that in such a case
$\VaR_{H,t,c}$ obeys the following equation:

\[ \int_0^{\infty} (1- F_{X,t,h}(-\VaR_{H,t,c}))d F_{H,t}(h) = c   \]

whereas $\ES_{H,t,c}$ is given by

\[ \ES_{H,t,c} = - \frac{1}{1-c} \int_0^{\infty}  \Ex[ X_{t+h} - X_t | X_{t+h} - X_t \le - \VaR_{H,t,c}]
\mbox{Prob}( X_{t+h} - X_t \le - \VaR_{H,t,c}) d F_{H,t}(h) \]

For the specific Gaussian case~(\ref{eq:normaldX}) we have

\[ \int_0^{\infty} \Phi\left(\frac{\mu_{t,h} + \VaR_{H,t,c}}{\sigma_{t,h}}   \right) d F_{H,t}(h) = c   \]
\[ \ES_{H,t,c} =\frac{1}{1-c} \int_0^\infty \left[-\mu_{t,h} \Phi\left(\frac{-\mu_{t,h} - \VaR_{H,t,c}}{\sigma_{t,h}}   \right)   + \sigma_{t,h} p\left(\frac{-\mu_{t,h} - \VaR_{H,t,c}}{\sigma_{t,h}}   \right)\right] d F_{H,t}(h)
\]

Notice that in general one can try and obtain the quantile $\VaR_{H,t,c}$ for the random horizon case by using a root search, and subsequently compute also the expected shortfall.
Careful numerical integration is needed to apply these formulas for general distributions of $H_t$. The case of Table \ref{fmr_bernoulli} is somewhat trivial, since in the case where  $H_0$ is as in Table \ref{shp_bernoulli} integrals reduce to summations of two terms.

We note also that the maximum difference, both in relative and absolute terms, between ES and VaR is reached by the model under random holding period $H_0$. Under this model the change in portfolio value shows heavier tails than under a single deterministic holding period. In order to explore the impact of SHP's distribution tails on the liquidity-adjusted risk, in the following we will simulate SHP models with $H_0$ distributed as an Exponential,  an Inverse Gamma distribution\footnote{obtained by rescaling a distribution IG$\left(\frac{\nu}{2},\frac{\nu}{2}\right)$ with $\nu=3$. Before rescaling, setting $\alpha=\nu/2$, the inverse gamma density is
 $f(x) = (1/\Gamma(\alpha)) (\alpha)^{\alpha} x^{-\alpha - 1} e^{-\alpha/x}$, $x > 0$, $\alpha >0$, with expected value $\alpha/(\alpha - 1)$. We rescale this distribution by $k = 8.66/(\alpha/(\alpha - 1))$ and take for $H_0$ the random variable with density $f(x/k)/k$} and a Generalized Pareto distribution\footnote{with scale parameter $k =9$ and shape parameter $\alpha = 2.0651$, with cumulative distribution function $F(x)=1-\left(\frac{k}{k+x} \right )^\alpha$, $x \ge 0$, this distribution has moments up to order $\alpha$. So the smaller $\alpha$, the fatter the tails. The mean is, if $\alpha > 1$, $\Ex[H_{0}] = k/(\alpha-1)$} having parameters calibrated in order to obtain a sample with the same 99\%-quantile of 75 business days:

\begin{table}[H]
\caption{SHP statistics and resulting Market Risk}
\centering
\begin{tabular}{c c c c|c c c c c }
  \hline\hline
  Distribution & Mean & Median & 99\%-q   & VaR 99.96\% & VaR 99.96\% & ES 99.96\% & ES 99.96\% & ES/VaR-1\\[0.5ex]
               &      &        &               & simulation  & root search  &  simulation & root search &  \\[0.5ex]
  \hline
  Exponential & 16.3 & 11.3 & 75.0 & 39.0 & 39.2  & 44.7  &  44.7 & 14 \%\\
  Pareto  & 8.45 & 3.7 & 75 & 41.9  & 41.9    &  57.1   &  56.9  & 36\%   \\[1ex]
  Inverse Gamma & 8.6 & 3.7 & 75.0 & 46.0 & 46.7  & 73.5  &  73.0  &  55 \%  \\[1ex]
  \hline
\end{tabular}
\label{fmr_various}
\end{table}

The SHP process changes the statistical nature of the P\&L process: the heavier the tails of the SHP distribution, the heavier the tails of P\&L distribution. Notice that our Pareto distribution has tails going to 0 at infinity with exponent around $3$, as one can see immediately by differentiation of the cumulative distribution function,  whereas our inverse gamma has tails going to 0 at infinity with exponent about 2.5. In this example we have that the tails of the inverse gamma are heavier, and indeed for that distribution VaR and ES are larger and differ from each other more. This can change of course if we take different parameters in the two distributions.

\section{Dependence modelling: a bivariate case}

Within multivariate modelling, using a common SHP for many normally distributed risks leads to dynamical versions of the so-called {\it normal mixtures} and {\it normal mean-variance mixtures} \cite{qrm}.

Let the log-returns (recall $X^i_t = \ln V^i_t$, with $V^i_t$ the value at time $t$ of the i-th asset)
\[ X^1_{t+h} - X^1_t, \ldots, X^m_{t+h} - X^m_t \]
be normals with means $\mu^1_{t,h}, \ldots, \mu^m_{t,h}$ and covariance matrix $Q_{t,h}$.

Then
\[
P[X^1_{t+H_t} - X^1_t < x_1, \ \ X^m_{t+H_t} - X^m_t< x_m] = \int_0^\infty
P[X^1_{t+h} - X^1_t < x_1, \ \ X^m_{t+h} - X^m_t< x_m] d F_{H,t}(h)
 \]
is distributed as a mixture of multivariate normals,
and a portfolio $V_t$ of the assets $1,2,...,m$ whose log-returns $X_{t+h}-X_t$ $(X_t = \ln V_t)$ are a linear weighted combination
$w_1,...,w_m$ of the single asset log-returns $X^i_{t+h}-X^i_t$  would be distributed as
\[ P [ X_{t+H_t} - X_t < z ] =
\int_0^\infty
P[w_1(X^1_{t+h} - X^1_t) + \ldots + w_m(X^m_{t+h} - X^m_t) < z]  d F_{H,t}(h)
\]

%
%
%

In particular, in analogy with the unidimensional case, the mixture
may potentially generate skewed and fat-tailed distributions, but when working with more than one
asset this has the further implication that VaR is not guaranteed to
be subadditive on the portfolio. Then the risk manager who wants to
take into account SHP in such a setting should adopt a coherent
measure like Expected Shortfall.

A natural question at this stage is whether the adoption of a common SHP can add dependence to returns that are jointly Gaussian under deterministic calendar time, perhaps to the point of making extreme scenarios on the joint values of the random variables possible.

Before answering this question, one needs to distinguish extreme behaviour in the single variables and in their joint action in a multivariate setting.
Extreme behaviour on the single variables is modeled for example by heavy tails in the marginal distributions of the single variables.
Extreme behaviour in the dependence structure of say two random variables is achieved when the two random variables tend to take extreme values in the same direction together. This is called tail dependence, and one can have both upper tail dependence and lower tail dependence. More precisely, but still loosely speaking, tail dependence expresses the limiting proportion according to which the first variable exceeds a certain level given that the second variable has already exceeded that level. Tail dependence is technically defined through a limit, so that it is an asymptotic notion of dependence. For a formal definition we refer, for example, to \cite{qrm}. ``Finite" dependence, as opposed to tail, between two random variables is best expressed by rank correlation measures such as Kendall's tau or Spearman's rho.

We discuss tail dependence first.
In case the returns of the portfolio assets are jointly Gaussian
with correlations smaller than one, the adoption of a common random
holding period for all assets does not add tail dependence,
\textit{unless the commonly adopted random holding period has a
distribution with power tails}. Hence if we want to rely on one of
the random holding period distributions in our examples above to
introduce upper and lower tail dependence in a multivariate
distribution for the assets returns, we need to adopt a common
random holding period for all assets that is Pareto or Inverse Gamma
distributed. Exponentials, Lognormals or discrete Bernoulli
distributions would not work.  This can be seen to follow for
example from properties of the normal variance-mixture model, see
for example \cite{qrm}, page 212 and also Section 7.3.3.

A more specific theorem that fits our setup is Theorem 5.3.8 in
\cite{prestele}. We can write it as follows with our notation.

\begin{proposition}\label{prop:power} {\bf A common random holding period with less than power tails does not add tail dependence to jointly Gaussian returns.}
Assume the log-returns to be $W^i_t = \ln V^i_t$, with $V^i_t$ the value at time $t$ of the i-th asset, $i=1,2$, where
\[ W^1_{t+h} - W^1_t, W^2_{t+h} - W^2_t \]
are two correlated Brownian motions, i.e. normals with zero means, variances $h$ and instantaneous correlation less than 1 in absolute value:
\[ d \langle W^1, W^2 \rangle_t = d W^1_t \ d W^2_t = \rho_{1,2} dt,  \ \ |\rho_{1,2}| < 1. \]
Then adding a common non-negative random holding period $H_0$ independent of $W$'s leads to tail dependence in the returns
\[W^1_{H_0},W^2_{H_0} \]
if and only if $\sqrt{H_0}$ is regularly varing at $\infty$ with index $\alpha > 0$.
\end{proposition}

Theorem 5.3.8 in  \cite{prestele} also reports an expression for the tail dependence coefficients as functions of $\alpha$ and of the survival function of the student $t$ distribution with $\alpha + 1$ degrees of freedom.

Summarizing, if we work with power tails, the heavier are the tails of the common
holding period process $H$, the more one may expect {\it tail dependence} to emerge
for the multivariate distribution: by adopting a common SHP for all
risks, dependence could potentially appear in the whole
dynamics, in agreement with the fact that liquidity risk is a
systemic risk.

We now turn to finite dependence, as opposed to tail dependence. First we note the well known elementary but important fact that one can have two random variables with very high dependence but without tail dependence. Or one can have two random variables with tail dependence but small finite dependence. For example, if we take two jointly Gaussian Random variables with correlation 0.999999, they are clearly quite dependent on each other but they will not have tail dependence, even if a rank correlation measure such as Kendall's $\tau$ would be $0.999$, still very close to $1$, characteristic of the co-monotonic case. This is a case with zero tail dependence but very high finite dependence.
On the other hand, take a bivariate student $t$ distribution with few degrees of freedom and correlation parameter $\rho = 0.1$. In this case the two random variables have positive tail dependence and it is known that Kendall's tau for the two random variables is
\[ \tau = \frac{2}{\pi} \arcsin(\rho) \approx 0.1 \]
which is the same tau one would get for two standard jointly Gaussian random variables with correlation $\rho$. This tau is quite low, showing that one can have positive tail dependence while having very small finite dependence.

The above examples point out that one has to be careful in distinguishing large finite dependence and tail dependence.

A further point of interest in the above examples comes from the fact that the multivariate student $t$ distribution can be obtained by the multivariate Gaussian distribution when adopting a random holding period given by an inverse gamma distribution (power tails).  We deduce the important fact that in this case {\emph{a common random holding period with power tails adds positive tail dependence but not finite dependence}}.

In fact, one can prove a more general result easily by resorting to the tower property of conditional expectation and from the definition of tau based on independent copies of the bivariate random vector whose dependence is being measured. One has the following ''no go" theorem for increasing Kendall's tau of jointly Gaussian returns through common random holding periods, regardless of the tails power.

\begin{proposition}{\bf A common random holding period does not alter Kendall's tau for jointly Gaussian returns.}
Assumptions as in Proposition \ref{prop:power} above.
Then adding a common non-negative random holding period $H_0$ independent of $W$'s leads to the same Kendall's tau for
\[W^1_{H_0},W^2_{H_0} \]
as for the two returns
\[W^1_{t},W^2_{t} \]
for a given deterministic time horizon $t$.
\end{proposition}

Summing up, this result points out that adding further
finite dependence through common SHP's, at least as measured by Kendall's tau, can be impossible if we start
from Gaussian returns. A different popular rank correlation measure, Spearman's rho, does not
coincide for the bivariate $t$ and Gaussian cases though, so that it is not excluded that
dependence could be added in principle though dependent holding periods, at least if we measured dependence with
Spearman's $\rho$. This is under investigation.


More generally, at least from a theoretical point of view, it could be
interesting to model other kinds of dependence than the one stemming
purely from a common holding period (with power tails). One could have
two different holding periods that are themselves dependent on each
other in a less simplistic way, rather than being just identical.
In this case it would be interesting to study the tail dependence implications and also finite dependence as measured by Spearman's rho.

We will investigate this aspect in further research, but increasing dependence may
require, besides the adoption of power tail laws for the random
holding periods, abandoning the Gaussian distribution for the basic
assets under deterministic calendar time.

A further aspect worth investigating is the possibility to calculate semi-closed form risk contributions to VaR and ES under SHP along the lines suggested in \cite{tasche06}, and to investigate the Euler principle as in \cite{tasche08} and \cite{tasche09}.

\section{Calibration over liquidity data}

We are aware that multivariate SHP modelling is a purely theoretical
exercise and that we just hinted at possible initial developments
above. Nonetheless, a lot of financial data is being collected by
regulators, providers and rating agencies, together with a
consistent effort on theoretical and statistical studies. This will
possibly result in available synthetic indices of liquidity risk
grouped by region, market, instrument type, etc. For instance, Fitch
already calculates market liquidity indices on CDS markets
worldwide, on the basis of a scoring proprietary model.

\paragraph{Dependences between liquidity, credit and market risk}
It could be an interesting exercise to calibrate the dependence
structure (e.g. copula function) between a liquidity index (like the
Fitch's one), a credit index (like iTRAXX) and a market index (for
instance Eurostoxx50) in order to measure the possible (non linear)
dependence between the three. The risk manager of a bank could use
the resulting dependence structure within the context of risk
integration, in order to simulate a joint dynamics as a first step,
to estimate later on the whole liquidity-adjusted VaR/ES by assuming
co-monotonicity between the variations of the liquidity index and of
the SHP processes.

\paragraph{Marginal distributions of SHPs} A lot of information on SHP
`extreme' statistics of a OTC derivatives portfolio could be
collected from the statistics, across Lehman's counterparties, of
the time lags between the Lehman's Default Event Date and the trade
dates of any replacement transaction. The data could give
information on the marginal distribution of the SHP of a portfolio,
in a stressed scenario, by assuming a statistical equivalence
between data collected `through the space' (across Lehman's
counterparties) and `through the time' under i.i.d.
hypothesis\footnote{a similar approach is adopted in \cite{or}
within the context of operational risk modelling}. The risk manager
of a bank could examine a more specific and non-distressed dataset
by collecting information on the ordinary operations of the business
units.

\section{Conclusions}
Within the context of risk integration, in order to include
liquidity risk in the whole portfolio risk measures, a stochastic
holding period (SHP) model can be useful, being versatile, easy to
simulate, and easy to understand in its inputs and outputs. In a
single-portfolio framework, as a consequence of introducing a SHP
model, the statistical distribution of P\&L moves to possibly heavier tailed
and skewed mixture distributions\footnote{even if the originary return process was
Brownian, implying normal returns, the resulting P\&L process under
SHP is not: paraphrasing Geman {\it et al.} \cite{apb}, it is not
the time for a process to be Brownian!}. In a multivariate setting,
the dependence among the SHP processes to which marginal P\&L are
subordinated, may lead to dependence on the latter under drastic choices of the SHP distribution,
and in general to
heavier tails on the total P\&L distribution. At present, lack of
synthetic and consensually representative data forces to a
qualitative top-down approach, but it is straightforward to assume
that this limit will be overcome in the near future.

\end{document}